\documentclass[12pt]{iopart}
\usepackage{iopams} 
\usepackage[latin1]{inputenc}
\usepackage{graphicx}
\usepackage{url}

\begin{document} 
 
\title[Spontaneous emission -- graphene waveguides -- surface plasmon polaritons]{Surface plasmon enhancement of spontaneous emission in graphene waveguides}
\author{Mauro Cuevas} 
\address{Consejo Nacional de
Investigaciones Cient\'ificas y T\'ecnicas (CONICET) and Facultad de Ingenier\'ia y Tecnolog\'ia Inform\'atica, Universidad de Belgrano,
Villanueva 1324, C1426BMJ, Buenos Aires, Argentina}
\address{Grupo de Electromagnetismo Aplicado, Departamento de F\'isica, FCEN, Universidad de Buenos Aires,  Ciudad Universitaria,
Pabell\'on I, C1428EHA, Buenos Aires, Argentina}
\ead{cuevas@df.uba.ar}

\begin{abstract} 
This work analyzes the spontaneous emission of a single emitter placed near the graphene waveguide formed by two parallel graphene monolayers, with an insulator spacer layer. In this case, the eigenmodes supported by the structure, such as surface plasmon and wave guided modes, provide decay channels for the electric dipole placed close to the waveguide. We calculated the contribution to the decay rate of symmetric and antisymmetric eigenmodes as a function of frequency and the orientation of the emitter.    Our results show that the modification of the spontaneous emission due to excitation of guided modes is much lower than the corresponding decays through the excitation of symmetric and antisymmetric surface plasmons, for which, the spontaneous emission is dramatically enhanced. As a consequence of the high confinement of surface plasmons in the graphene  waveguide, we found that the decay rate of the emitter with  vertical  orientation (with respect to graphene sheets) is twice the corresponding decay of the same emitter with parallel  orientation  in the whole frequency range where surface plasmon modes exist. Differently from metallo--dielectric structures,  where structural parameters determine the range and magnitude of this emission, our work shows that, by dynamically tuning the chemical potential of graphene,  the spectral region where the decay rate is enhanced can be chosen over a wide range.
\end{abstract} 



\maketitle

\section{Introduction} 

A property of materials that exhibit a real negative electric permittivity --such as metals-- is their capacity to guide surface plasmons (SPs) along their boundary 
\cite{raether,maier}. SPs propagate along the surface with a periodicity lower than the wavelength of same frequency electromagnetic radiation, which is a suitable feature for the miniaturization of photonic devices. In bounded geometries, these modes, called localized surface plasmons, are characterized by discrete frequencies that depend on the size and shape of the object to which they are confined. The  localization provided by SPs is quite adequate for many applications such as data storage, microscopy, light generation, biochemical sensing, antennas working at nanoscale \cite{barnes,maier_nature} and control applications like light trapping \cite{guo}.

Apart from the well known SPs supported by an insulator--metal interface, long livid SPs can be supported by graphene --a 2 D sheet of carbon atoms arranged in a honeycomb lattice \cite{novoselov}--  from terahertz up to mid--infrared frequencies \cite{jablan,rana}. High confinement, relative low loss, and good tunability of surface plasmon spectrum through electrical or chemical modification of the carrier densitiy \cite{jablan,a_primer,garcia_de_abajo,LSP,RCD},  are three characteristics that make graphene a promising plasmonic alternative material to noble metals.

This paper deals with the modification of the spontaneous emission of a single optical emitter by interaction with its local enviroment \cite{purcell}, a process that plays a key role in the realization of current light control devices, such as photonic band gaps \cite{CP}, high efficient single photon sources \cite{SPS2} or single--photon transistors \cite{spt}.   
 Since SPs are non radiative modes trapped on the surface, an emitter close to a plane surface can be regarded as an element of surface roughness that serves to couple photons to SPs \cite{philpott}.  Experiments concerning this coupling have proved the non radiative energy transfer to SPs on a planar metallic interface \cite{weber,pockrand}.

Rigorous classical theoretical approach, using Hertz--vector representation of electromagnetic fields \cite{chance,lukos} or Green's tensor approach \cite{chance2} have been applied to determine the contribution of the radiating and evanescent modes to the power emitted by a source close to the reflecting surface. The coupling between a single emitter and metals surface states was also studied in the framework of quantum electrodynamics, as emission stimulated by zero--point fluctuations of the electromagnetic field \cite{philpott,sipe,agarwal,yablonovich}. In the limit of weak coupling, results of quantum mechanical calculations have been found to be similar to those derived by classical electromagnetic theory \cite{philpott,sipe,novotny,greffet2}. Both classical and quantum formalisms have been applied to study the spontaneous emission of a single emitter near a planar microcavity characterized by more than two interfaces. In these structures, wave guided modes (WG) are resonant optical modes that, like SPs, provide new channels for the spontaneous decay rate of a single emitter \cite{barnes1,barnes2,josab19,pra84}. Enhanced emission rate due to excitation of $p$ and $s$ polarized SPs or WG modes on negative index material multilayers has been reported \cite{LHM1,LHM2,LHM3}. A variety of structures such as
uniform planar microcavities \cite{tunable1},  periodically patterned metallic or dielectric membranes  (2--D photonic crystal) \cite{yablonovich2,vukovic}, cylindrical nanowires \cite{wire1} and gratings \cite{barnes3} have been the object of intensive research over the last few years due to the  possibility to engineer the WG  or SP mode density of states and consequently modify the emission into a particular mode. 

Interactions between single optical emitters and SPs on graphene have been investigated in different structures, such as infinite graphene monolayers \cite{grafeno,grafeno1,grafeno2,grafeno4}, ribbons or nanometer sized disks \cite{grafeno3}.  Double--layer graphene waveguides have also become the focus of particular attention. 
For example, a thin glass film coated with graphene and with a dipole emitter embedded at the center of the glass has been recently proposed  \cite{guia2}.  On the other hand, coupling between a single emitter and SPs in paired graphene layers has recently been examined \cite{guia1}. To efficiently couple the fundamental SP mode, the emitter is set to be vertically polarized to graphene layers and positioned at the center of the gap between them. However, there is no reference in the  literature about a comprehensive examination of the role played by each of  the waveguide SPs in modifying the spontaneous emission rate for  an arbitrary polarization of the emitter.

In this paper we analytically study the spontaneous emission rate of a dipole located above a waveguide formed by two parallel graphene sheets with an insulator spacer layer, and we present results showing the role of the  eigenmodes of the structure (SPs or WG modes) in modifying the spontaneous emission rate with respect to the rate in absence of the waveguide. One of the interesting  differences with a  single  graphene monolayer structure is  that the graphene waveguide studied here has two  graphene interfaces, each of which may carry SP modes, and the fields of these modes can overlap through the gap dielectric layer, leading SPs into separated branches. By exploiting the separation between the two graphene sheets as a degree of freedom, it is possible to modify these branches and consequently  their influence on the spontaneous emission rate. In this context, several works  focused on the role that the eigenmodes play on metallic waveguides \cite{pra84,tunable1,prb85} or  on metamaterial waveguides \cite{LHM2,LHM3,pra78}. 
In addition, when graphene is included, by exploiting the chemical potential on graphene monolayers as another degree of freedom, one can  
shift 
these properties to other frequency regions.   

The plan of the paper is as follows. In section \ref{teoria}, we sketch the classical electromagnetic formalism  based on the calculation of the electric vector potential.  By virtue of the translational invariance of the system along a  plane parallel to graphene sheets ($x-y$ plane), we reduce the  solution of the original vectorial problem to the treatment of two scalar problems corresponding to the basic modes of polarization $p$ (magnetic field parallel to the $x-y$ plane) and  $s$ (electric field parallel  to the $x-y$ plane).    
In section \ref{SE} we provide a general expression for the spontaneous decay rate of an  oscillating emitter with an arbitrary orientation of its dipole moment. 
Assuming that the graphene surface conductivity follows the Kubo model, we determine the eigenmode dispersion curves -- that is, the real and the imaginary parts of the eigenmode propagation constant as  functions of the frequency -- and we present approximated analytic expressions for the spontaneous decay rate into these eigenmodes.
 By applying the residues method, in section \ref{resultados} we calculate the contribution of each eigenmode to the total decay rate. We find that, the decay rate near the interface through SPs is much larger (by over five orders of magnitude) than the decay rate through guided modes. 
Within  the framework of quantum electrodynamics, results are discussed in terms of both the surface plasmon density of states and  the effective mode length.
 Concluding remarks are provided in Section \ref{conclusiones}. The Gaussian system of units is used
 and an $\mbox{exp}(-i\, \omega\, t)$ time--dependence is implicit throughout the paper, with $\omega$ as the angular frequency, $t$ as the time, and $i=\sqrt{-1}$. The symbols Re and Im are used for denoting the real and imaginary parts of a complex quantity, respectively.

\section{Electromagnetic field of a radiating dipole} \label{teoria} 

\begin{figure}[htbp]
\centering
\resizebox{0.50\textwidth}{!}
{\includegraphics{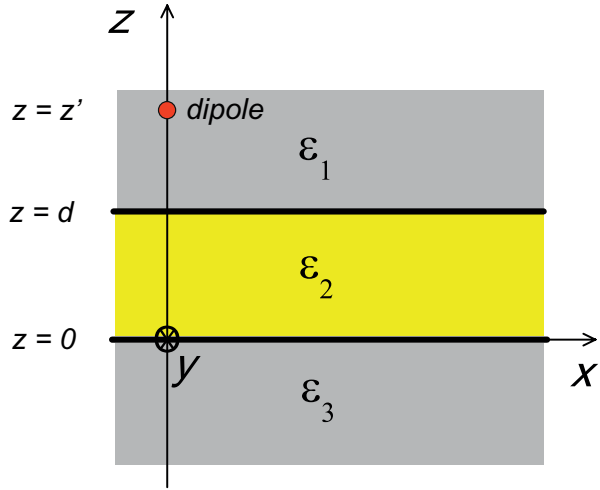}}
\caption{\label{fig:epsart} Schematic illustration of the system. Two graphene sheets, characterized by surface conductivity $\sigma$, are embedded between adjacent dielectric layers at $z=0$ and $z=d$. The electric dipole is located at $\vec{x}=z^{'}\hat{z}$.}\label{sistema}
\end{figure}

Let us consider a structure made up of three linear, isotropic and homogeneous media arranged as shown in Figure \ref{sistema}. The interfaces of the layers are parallel to the $x-y$ plane. It is assumed that the graphene monolayers are embedded between adjacent dielectric layers, 
at $z=0$ and $z=d$. An electric dipole is located at $\vec{x}=z^{'} \hat{z}$, at a distance $z^{'}>d$ from the plane interface  $z=0$. The current density of the electric dipole with moment $\vec{p}$ placed at $\vec{x}=\vec{x^{'}}$ is
\begin{eqnarray}\label{corrienteje}
\vec{j_e}(\vec{x})=-i \omega \vec{p} \, \delta(\vec{x}-\vec{x}').
\end{eqnarray} 
Since the decay rate can be related to the electric field induced by a dipole itself, the aim of this section is to derive  an analytical expression for the scattered electric field  in the same region where the dipole is embedded. Taking into account the infinitesimal translational invariance in the $x$ and $y$ directions, 
the  field of the electric dipole can be represented as a superposition of two  basic polarization modes: $p$ polarization mode, for which the magnetic field  is parallel to the $x-y$ plane in Figure \ref{sistema}, and $s$ polarization mode, for which the electric field is parallel to the  $x-y$ plane. From the mathematical point of view,  the electromagnetic field can be represented by two scalar functions $a_p(\vec{x})$ and $a_s(\vec{x})$ which are, respectively, the $z$ component electric and magnetic vector potentials \cite{lukos,novotny},
\begin{eqnarray}
\vec{A_e}(\vec{x})=a_p(\vec{x})\,\hat{\textbf{z}}, \nonumber \\
\vec{A_h}(\vec{x})=a_s(\vec{x})\,\hat{\textbf{z}}. \label{ApAs}
\end{eqnarray}
The electric field $\vec{E}=\vec{E}_p(\vec{x})+\vec{E}_s(\vec{x})$ and the magnetic field $\vec{H}=\vec{H}_p(\vec{x})+\vec{H}_s(\vec{x})$ can be derived according to
\begin{eqnarray}\label{camposEHp}
\begin{array}{cc}
\vec{E}_p(\vec{x})= i k_0 \mu  \left[
\begin{array}{c}
\frac{1}{k^2} \frac{\partial }{\partial z}\frac{\partial a_p}{\partial x} \\

\frac{1}{k^2} \frac{\partial }{\partial z} \frac{\partial a_p}{\partial y}\\

(1+\frac{\partial^2 }{\partial z^2}) a_p
\end{array}\right], &
\vec{H}_p(\vec{x})=  \left[
\begin{array}{c}
\frac{\partial a_p}{\partial y}\\
-\frac{\partial a_p}{\partial x}\\
0
\end{array}\right],
\end{array}
\end{eqnarray} 
\begin{eqnarray}\label{camposEHs}
\begin{array}{cc}
\vec{E}_s(\vec{x})=\left[
\begin{array}{c}
-\frac{\partial a_s}{\partial y}\\
\frac{\partial a_s}{\partial x}\\
0
\end{array}\right], &
\vec{H}_s(\vec{x})= i k_0 \varepsilon  \left[
\begin{array}{c}
\frac{1}{k^2} \frac{\partial }{\partial z}\frac{\partial a_s}{\partial x} \\

\frac{1}{k^2} \frac{\partial }{\partial z} \frac{\partial a_s}{\partial y}\\

(1+\frac{\partial^2}{\partial z^2}) a_s
\end{array}\right]. 
\end{array}
\end{eqnarray} 
The direct field emitted by the dipole placed at $\vec{x}=\vec{x^{'}}$  
is written as \cite{lukos,novotny,chew} 
\begin{equation}\label{Ae}
a_\tau(\vec{x})=\frac{k_0}{2 \pi} \int_{-\infty}^\infty \, \frac{1}{\gamma^{(1)}} d_\tau(\alpha,\beta) \,e^{i\gamma^{(1)|z-z^{'}|}} e^{i[\alpha x +\beta y]} d\alpha d\beta,
\end{equation}  
where $\tau=p,s$ indicates the polarization mode,  $\gamma^{(1)}=\sqrt{k_1^2-\alpha^2-\beta^2}$, $k_1^2=k_0^2\varepsilon_1 \mu_1$, $k_0=\omega/c$ is the modulus of the photon wave vector in vacuum, $\omega$ is the angular frequency, $c$ is the vacuum speed of light, $\varepsilon_1$ and $\mu_1$ are the electric permittivity and magnetic permeability, respectively, and  the spectral functions $d_\tau$ are  given by \cite{lukos,novotny}  
\begin{eqnarray}\label{pym}
\begin{array}{l}
d_s=\frac{k_1^2}{k_0 \varepsilon} \left[-\frac{\beta}{\alpha^2+\beta^2}p_x+\frac{\alpha}{\alpha^2+\beta^2}p_y\right] 
\\  
d_p^{\pm}=\mp\frac{\alpha \gamma^{(1)}}{\alpha^2+\beta^2}p_x\mp\frac{\beta \gamma^{(1)}}{\alpha^2+\beta^2}p_y+p_z.  
\end{array}
\end{eqnarray}
Note that in lossless media the quantities $\gamma^{(1)}$ are real or purely imaginary. In the first case, which occurs in the so--called radiative zone $\sqrt{\alpha^2+\beta^2}< (\omega/c)\sqrt{\varepsilon_1 \mu_1}$, the integrand in Eq. (\ref{potencialcapam}) represents plane waves propagating away from the dipole along a direction that forms an  angle $\theta$ ($\sin \theta = c / (\omega \sqrt{\varepsilon_1 \mu_1})$) with the $\pm z$ axis. In the second case, which occurs in the so--called non radiative zone $\sqrt{\alpha^2+\beta^2}> (\omega/c)\sqrt{\varepsilon_1 \mu_1}$, these fields represent evanescent waves that attenuate for $z\rightarrow \pm \infty$. 

The infinitesimal translational  invariance in $x$ and $y$ directions of the system for which  the scalar potentials are being searched, allows us to write the Fourier representation of scalar potentials $a_p(\vec{x})$ and $a_s(\vec{x})$  
like  
\begin{eqnarray}\label{potencialcapam}
\begin{array}{l}
a_\tau^{(m)}(\vec{x})=\frac{k_0}{2 \pi} \int_{-\infty}^{\infty} d\alpha\,d\beta \,f^{(m)}_\tau(\alpha,\beta,z,z^{'}) e^{i(\alpha x+\beta y)}
\end{array}
\end{eqnarray}
where functions  $f_\tau^{(m)}(\alpha,\beta,z,z^{'})$ ($m=1,\,2,\,3$) depend on the location of the source and of the polarization mode $\tau=p,s$.

The integrand in (\ref{potencialcapam}) is written as 
\begin{eqnarray}\label{fmedio1}
\begin{array}{l}
f_\tau^{(1)}(\alpha,\beta,z,z^{'})=\frac{1}{\gamma^{(1)}} d_\tau \,e^{i\gamma^{(1)|z-z^{'}|}}+A^{(1)}_\tau\,e^{i\gamma^{(M)}z}, 
\end{array}
\end{eqnarray}
%
\begin{eqnarray}\label{fmedio2}
\begin{array}{l}
f_\tau^{(2)}(\alpha,\beta,z,z^{'})= A^{(2)}_\tau\,e^{i\gamma^{(2)}z}+B^{(2)}_\tau\,e^{-i\gamma^{(2)}z},
\end{array}
\end{eqnarray}
%
\begin{eqnarray}\label{fmedio3}
\begin{array}{l}
f_\tau^{(3)}(\alpha,\beta,z,z^{'})= B^{(3)}_\tau\,e^{-i\gamma^{(3)}z},
\end{array}
\end{eqnarray}
%
where the superscript $m=1,\,2,\,3$ denotes medium 1 ($z>d$), medium 2 ($0<z<d$) or medium 3 ($z<0$), and $\gamma^{(m)}=\sqrt{k_m^2-(\alpha^2+\beta^2)}$, with $k_m^2=k_0^2\varepsilon_{m}\mu_{m}$, is the normal component of the wave vector in each homogeneous region. The complex coefficients $A^{(m)}_\tau$ and $B^{(m)}_\tau$  in Eqs. (\ref{fmedio1}) to (\ref{fmedio3})  correspond to   the amplitude of upgoing ($+z$ propagation direction) and downgoing ($-z$ propagation direction) plane waves, respectively, and they are  solutions of Helmholtz equation, whereas 
the former term in Eq. (\ref{fmedio1}) is  
associated to the primary dipole emission of the  source. 
There are two types of boundary conditions which must fulfill the solutions given by Eqs. (\ref{potencialcapam}) to (\ref{fmedio3}), boundary conditions at $z=\pm \infty$ and boundary conditions at interfaces $z=0$ and $z=d$. The former requires either outgoing waves at infinity  or exponentially decaying waves at infinity, depending on the values of $\alpha$, $\beta$ and $\omega$. 
The boundary conditions on interfaces   $z=0$ and $z=d$ impose that
\begin{eqnarray}
\hat{\textbf{z}}\times[\vec{E}^{(m)}-\vec{E}^{(m+1)}]|_{z=d_{m}}=0, \label{cc1} \nonumber\\ 
\hat{\textbf{z}}\times[\vec{H}^{(m)}_{m}-\vec{H}^{(m+1)}]|_{z=d_{m}} =\frac{4\pi \sigma}{c} \hat{\textbf{z}}\times \vec{E}|_{z=d_{m}} \label{cc2}
\end{eqnarray}
where $\sigma$ is the graphene conductivity, $d_1=d$ and $d_2=0$.  Inserting the expressions for  $\vec{E}_p$ and $\vec{H}_p$ given by  Eq. (\ref{camposEHp})  into Eqs. (\ref{cc2}) we obtain the following conditions on the scalar potential $a_p(\vec{x})$,  
\begin{eqnarray} 
\frac{1}{\varepsilon_{m}} \frac{\partial a_p^{(m)}}{\partial z}|_{z=d_{m}}=\frac{1}{\varepsilon_{m+1}} \frac{\partial a_p^{(m+1)}}{\partial z}|_{z=d_{m}}, \label{cc2p} \nonumber\\
a_p^{(m)}|_{z=d_{m}}-a_p^{(m+1)}|_{z=d_{m}}= 
\frac{4\pi  \sigma}{c} \frac{i}{k_0 \varepsilon_{m+1}} \frac{\partial a_p^{(m)}}{\partial z}|_{z=d_{m}}.   \label{cc1p}
\end{eqnarray}
Similarly, by using Eqs. (\ref{camposEHs}) and (\ref{cc2}) we obtain  following conditions on the scalar potential $a_s(\vec{x})$, 
\begin{eqnarray}
a_s^{(m)}|_{z=d_{m}}=a_s^{(m+1)}|_{z=d_{m}},   \label{cc2s} \nonumber\\
\frac{1}{\mu_{m}} \frac{\partial a_s^{(m)}}{\partial z}|_{z=d_{m}}-\frac{1}{\mu_{m+1}} \frac{\partial a_s^{(m+1)}}{\partial z}|_{z=d_{m}}=
-\frac{4\pi}{c} i k_0 a_s^{(m)}|_{z=d_{m}}. \label{cc1s}
\end{eqnarray}
To obtain the complex amplitudes  $A^{(m)}_\tau$ and $B^{(m)}_\tau$  
we must combine     Eq. (\ref{potencialcapam}), with $f_\tau^{(m)}$ given by Eqs. (\ref{fmedio1}) to (\ref{fmedio3}),  with conditions (\ref{cc2p}) and (\ref{cc2s})  for $\tau=p$ and  $\tau=s$ polarization, respectively. 
Here, we write  the amplitude corresponding to region $m=1$, where the dipole is placed,
\begin{eqnarray} \label{a1}
A^{(1)}_\tau=\frac{1}{\gamma^{1}}   
r_\tau^{(1,3)} e^{i\gamma^{(1)}(z^{'}-2d)} \, d_\tau^{-},
\end{eqnarray}  
where 
\begin{eqnarray} \label{a1b}
r_\tau^{(1,3)}=   
\frac{r^{(1,2)}_\tau +r^{(2,3)}_\tau F_\tau e^{i \gamma^{(2)} 2 d} }{1-r^{(2,1)}_\tau r^{(2,3)}_\tau e^{i \gamma^{(2)} 2 d} },  
\end{eqnarray} 
and
\begin{eqnarray}\label{F} 
F_\tau=t^{(1,2)}_\tau t^{(2,1)}_\tau-r^{(1,2)}_\tau r^{(2,1)}_\tau. 
\end{eqnarray}  
The complex amplitudes
\begin{eqnarray}\label{fresnelp}
r^{(i,j)}_p=\frac{\frac{\gamma^{(i)}}{\varepsilon_i}-\frac{\gamma^{(j)}}{\varepsilon_j}+\frac{4 \pi \sigma}{c k_0} \frac{\gamma^{(i)}}{\varepsilon_i} \frac{\gamma^{(j)}}{\varepsilon_j} }{\frac{\gamma^{(i)}}{\varepsilon_i}+\frac{\gamma^{(j)}}{\varepsilon_j}+\frac{4 \pi \sigma}{c k_0} \frac{\gamma^{(i)}}{\varepsilon_i} \frac{\gamma^{(j)}}{\varepsilon_j}},\\
t^{(i,j)}_p=\frac{2 \frac{\gamma^{(i)}}{\varepsilon_i} }{\frac{\gamma^{(i)}}{\varepsilon_i}+\frac{\gamma^{(j)}}{\varepsilon_j}+\frac{4 \pi \sigma}{c k_0} \frac{\gamma^{(i)}}{\varepsilon_i} \frac{\gamma^{(j)}}{\varepsilon_j}},
\end{eqnarray}
are the Fresnel reflection and transmission coefficients, respectively, for $p$ polarization, whereas 
\begin{eqnarray}\label{fresnels}
r^{(i,j)}_s=\frac{\frac{\gamma^{(i)}}{\mu_i}-\frac{\gamma^{(j)}}{\mu_j}-\frac{4 \pi k_0 \sigma}{c}  }{\frac{\gamma^{(i)}}{\mu_i}+\frac{\gamma^{(j)}}{\mu_j}+\frac{4 \pi \sigma k_0}{c} },
\\
t^{(i,j)}_s=\frac{2 \frac{\gamma^{(i)}}{\mu_i} }{\frac{\gamma^{(i)}}{\mu_i}+\frac{\gamma^{(j)}}{\mu_j}+\frac{4 \pi \sigma k_0}{c} },
\end{eqnarray}
are the Fresnel reflection and transmission coefficients, respectively, for $s$ polarization. Note that, in the case of $\sigma=0$, \textit{i.e.}, in the absence of current density induced in each graphene sheet,   
$F_\tau$ given by Eq. (\ref{F}) is equal to unity and then the coefficient (\ref{a1b}) converges to the well known  reflection coefficient of  three--layer medium without graphene \cite{chew}. 

The potential of the scattered field in the  medium $m=1$ can be obtained subtracting the first term in Eq. (\ref{fmedio1}) corresponding  to the primary dipole field, 
\begin{eqnarray}\label{fsmedio1}
\begin{array}{l}
f_\tau^{(1)}(\alpha,\beta,z,z^{'})|_{\,scatt}=f_\tau^{(1)}(\alpha,\beta,z,z^{'})-\\
\frac{1}{\gamma^{(1)}} d_\tau \,e^{i\gamma^{(1)|z-z^{'}|}}.
\end{array}
\end{eqnarray}
%
Introducing Eq. (\ref{fsmedio1}) into Eq. (\ref{potencialcapam}), and using Eqs. (\ref{camposEHp}) and (\ref{camposEHs}) we obtain an expression for the scattered electric field on region $z>d$ 
\begin{eqnarray}\label{campoE}
\vec{E}(\vec{x})|_{scatt}= \frac{i k_0}{2\pi}\int_{-\infty}^{+\infty} \Bigg\{ \left[
-\frac{k_0 \mu}{k_1^2}\alpha \gamma^{(1)}\hat{x}
-\frac{k_0 \mu}{k_1^2}\beta \gamma^{(1)}\hat{y}
+\frac{k_0 \mu}{k_1^2}(\alpha^2+\beta^2)\hat{z} 
 \right] A^{(1)}_p \nonumber \\ + 
\left[
-\beta \hat{x}
+\alpha \hat{y}
 \right] A^{(1)}_s \Bigg\} 
e^{i \gamma^{(1)}z} e^{i[\alpha x+\beta y]} d\alpha d\beta, 
\end{eqnarray}
where $A^{(1)}_p$ and  $A^{(1)}_s$ are given by Eq. (\ref{a1}).

\section{Spontaneous emission on a graphene waveguide}\label{SE}
The aim of this section is to derive a general formula of  the spontaneous decay  rate of an oscillating dipole placed above a graphene waveguide, 
paying special attention to the decay rate into the eigenmodes of the structure. All of the materials are non magnetic ($\mu_m=1, \,m=1,\,2,\,3$). The  waveguide is embedded in a transparent medium with an electric permittivity $\varepsilon_1=\varepsilon_3$ and the  region of space between  graphene sheets is filled with a transparent material with an  electric permittivity $\varepsilon_2$.

The graphene layer is considered as an infinitesimally thin, local and isotropic two--sided layer with frequency--dependent surface conductivity $\sigma(\omega)$ given by the Kubo formula \cite{falko,milkhailov}, which can be read as  $\sigma= \sigma^{intra}+\sigma^{inter}$, with the intraband and interband contributions being
\begin{equation} \label{intra}
\sigma^{intra}(\omega)= \frac{2i e^2 k_B T}{\pi \hbar (\omega+i\gamma_c)} \mbox{ln}\left[2 \mbox{cosh}(\mu_c/2 k_B T)\right],
\end{equation}  
\begin{eqnarray} \label{inter}
\sigma^{inter}(\omega)= \frac{e^2}{\hbar} \Bigg\{   \frac{1}{2}+\frac{1}{\pi}\mbox{arctan}\left[(\omega-2\mu_c)/2k_BT\right]-\nonumber \\
   \frac{i}{2\pi}\mbox{ln}\left[\frac{(\omega+2\mu_c)^2}{(\omega-2\mu_c)^2+(2k_BT)^2}\right] \Bigg\},
\end{eqnarray}  
where $\mu_c$ is the chemical potential (controlled with the help of a gate voltage), $\gamma_c$ the carriers scattering rate, $e$ the electron charge, $k_B$ the Boltzmann constant and $\hbar$ the reduced Planck constant.

\subsection{Radiated power}\label{radiacion}

According to Poynting theorem, the  time--averaged  radiated power $P$  by a dipole with a harmonic time dependence is given by \cite{jackson}
\begin{eqnarray}\label{Pradiada}
P=-\frac{1}{2}\int_V\,\mbox{Re}\left\{\vec{j_e}^*\cdot\vec{E}\right\}\,dV 
\end{eqnarray}  
where $V$  encloses   the source and $\vec{j}$ represents the source density current. Introducing the value of the current in   Eq. (\ref{corrienteje}), we obtain
\begin{eqnarray}\label{Pradiada2}
P=\frac{\omega}{2}\mbox{Im}\left\{\vec{p}^* \cdot \vec{E}(\vec{x^{'}}) \right\}
\end{eqnarray}  
where the field $\vec{E}$ is evaluated at the dipole position $\vec{x}'$. For an electric dipole above the plane waveguide interface we have
\begin{eqnarray}\label{E0Es}
\vec{E}(\vec{x})=\vec{E_0}(\vec{x})+\vec{E}(\vec{x})|_{scatt}
\end{eqnarray}  
where $\vec{E}_0(\vec{x})$ and $\vec{E}(\vec{x})|_{scatt}$ are the primary dipole field and the scattered field, respectively.  
Inserting Eq. (\ref{E0Es}) into Eq. (\ref{Pradiada2}) 
we obtain the radiated power  normalized with respect to the rate in absence of the waveguide  \cite{lukos,novotny}
\begin{eqnarray}\label{P_P0}
\frac{P}{P_0}=1+\frac{ \mbox{Im}\left\{ \vec{p}^* \cdot \vec{E}(\vec{x^{'}})|_{scatt} \right\}} {\mbox{Im}\left\{ \vec{p}^* \cdot \vec{E}_0(\vec{x^{'}})\right\}}=
1+\frac{3\,\varepsilon_1}{2p^2\,k_1^3} \,\mbox{Im}\left\{ \vec{p}^* \cdot \vec{E}(\vec{x^{'}})|_{scatt} \right\}.
\end{eqnarray}  
where $P_0=\omega\,p^2\,k_1^3/(3\,\varepsilon_1)$ is the total power  radiated by an electric dipole in the unbounded medium 1 \cite{novotny}.
Introducing the value of the electric field (\ref{campoE}), we obtain
\begin{eqnarray}\label{P_P0_1}
\frac{P}{P_0}=\cos^2\theta\,\left[\frac{P}{P_0}\right]_\bot+\sin^2\theta\,\left[\frac{P}{P_0}\right]_{||},
\end{eqnarray}  
where subscripts $\bot$ and $||$ indicate the normal and parallel orientation of the  dipole with respect to the $x-y$ plane, respectively, and       
\begin{eqnarray}\label{PP}
\left[\frac{P}{P_0}\right]_{\bot}= 
 1+\frac{3}{2 k_1^3}\, \mbox{Re} \int_0^{+\infty} dk_{||}\,\frac{k_{||}^3}{\gamma^{(1)}}\,
r_p^{(1,3)}(k_{||})\, e^{i2\gamma^{(1)}[z^{'}-d]}, 
\end{eqnarray}
%

\begin{eqnarray}\label{PP_paralelo}
\left[\frac{P}{P_0}\right]_{||}=
 1+\frac{3}{4 k_1^3}\, \mbox{Re}  \int_0^{+\infty} dk_{||}\,\frac{k_{||}}{\gamma^{(1)}}\,
[k_1^2\,r_s^{(1,3)}(k_{||})- \nonumber\\
\gamma^{(1)2} \, r_p^{(1,3)}(k_{||})]\, e^{i2\gamma^{(1)}[z^{'}-d]}, 
\end{eqnarray}
where the substitutions $\alpha=k_{||}\cos \phi_k$ and $\beta=k_{||}\sin \phi_k$ have been made, and $r_\tau^{(1,3)}$ is the reflection coefficient of  three--layer medium given by Eq. (\ref{a1b}), with $\varepsilon_1=\varepsilon_3$. 

The first term in  expressions (\ref{PP}) and (\ref{PP_paralelo}), equal to unity, corresponds to the direct dipole radiation in the homogeneous medium $1$. The integration range $[0,\,+\infty]$ can be divided into the two ranges $[0,\,k_1]$ and $[k_1,\,+\infty]$.  In the first range, the $z$--component of the  propagation wave vector $\gamma^{(1)}$ is real, which means the waves in medium 1 are propagating. This integral yields  the contribution  of multiple reflections on the graphene waveguide of  all the  plane waves emitted by the dipole at $\vec{x}=z^{'}\hat{z}$ and arriving at this position. In the  second range of integration $[0,\,+\infty]$, the $z$--component of the propagation wave vector $\gamma^{(1)}$ is imaginary, which means the waves in medium 1 are exponentially decaying  in the normal direction. This integral yields the contribution of the evanescent field radiated by the dipole, and thus it has a noticeable effect when the dipole is close enough to the interface $1$--$2$. 

\subsection{\label{plasmons} Decay through graphene eigenmodes}

Eigenmodes, like WG modes or SPs, may provide  decay channels for the electric dipole placed close to  the waveguide \cite{barnes1}.  The WG modes refer to modes which are evanescent waves in the two semi infinite regions (regions 1 and 3)  and standing waves in the insulator spacer layer (region 2), and SPs refer to  modes which propagate along the waveguide with their electric and magnetic fields decaying exponentially away from the graphene sheets  in all three regions.

As in any resonance phenomenon, the full characteristics of the electromagnetic eigenmodes supported by the graphene waveguide  can be obtained  by studying the singularities of the analytic continuation of the field (\ref{campoE}). 
Pole singularities occur at generally complex locations ($k_{||}$ is a complex magnitude)  and they represent  the propagation constant of the  eigenmodes supported by the graphene  waveguide. In the present case of the symmetric waveguide ($\varepsilon_1=\varepsilon_3$ and both graphene sheets with the same value of  the conductivity $\sigma$),  the dispersion equation  of $p$--polarized surface plasmons  splits into two branches \cite{gan}. 
The posibility of tuning the electronic properties of graphene  by adjusting the bias voltage  leads to unprecedented control over the location  of  plasmon resonances, for which this system has been suggested      as an  efficient plasmonic modulator \cite{svintsov}. Apart from these plasmon modes, $p$ and $s$ polarized guided modes   
can also be supported by the symmetric waveguide \cite{buslaev}.  

In order to obtain all the propagation characteristics of these eigenmodes,  the propagation constants are obtained by requiring the denominator in Eq. (\ref{a1b}) to be zero,
\begin{equation} \label{dispersion}
1-\left[r^{(2,1)}_\tau\right]^2  e^{i \gamma^{(2)} 2 d}  = 0,
\end{equation}  
where we have taken into account the equality  $\varepsilon_1=\varepsilon_3$. Physically, resonant condition (\ref{dispersion})  implies that a self--consistent field is established by means of a wave bouncing between the two boundaries of the layer at $z=0$ and $z=d$. This mean that the wave, after reflecting from the top and the bottom interfaces, together with a phase shift through the layer, should become in phase with itself again \cite{chew}. Because the expression for the field (\ref{campoE})  are derived from Maxwell's equations and boundary conditions, the resonant condition (\ref{dispersion}) holds for both homogeneous and inhomogeneous plane wave \cite{tamir}.
As a consequence of the waveguide symmetry, the modal fields are either odd or even with respect to the mirror symmetry  plane at  $z=d/2$. Following  the same procedure developed in \cite{gan,buslaev}, we can see that, for $p$ polarization Eq. (\ref{dispersion}) splits into two branches, one with a symmetric and the other with an antisymmetric magnetic field across the gap dieliectric layer
\begin{eqnarray}\label{modop}
\begin{array}{cc}
i \, \mbox{tan}(\gamma^{(2)}\frac{d}{2})=\frac{\gamma^{(1)} \varepsilon_2}{\gamma^{(2)} \varepsilon_1} \frac{1}{\left(1+\frac{4\pi \sigma \gamma^{(1)}}{\omega \varepsilon_1}\right)}    & H\mbox{--symmetric},   \\
i \, \mbox{tan}(\gamma^{(2)}\frac{d}{2})=\frac{\gamma^{(2)} \varepsilon_1}{\gamma^{(1)} \varepsilon_2} \left(1+\frac{4\pi \sigma \gamma^{(1)}}{\omega \varepsilon_1}\right)                & H\mbox{--antisymmetric}.
\end{array}
\end{eqnarray}  
Similarly, for $s$ polarization Eq. (\ref{dispersion}) splits into two branches, one with a symmetric and the other with an antisymmetric electric field across the gap dieliectric layer
\begin{eqnarray}\label{modos}
\begin{array}{cc}
i \, \mbox{tan}(\gamma^{(2)}\frac{d}{2})=\frac{\frac{\gamma^{(1)}}{\varepsilon_1}+\frac{4\pi \sigma \omega}{c^2}}{\gamma^{(2)} /\varepsilon_2}                & E\mbox{--symmetric},   \\
i \, \mbox{tan}(\gamma^{(2)}\frac{d}{2})=\frac{\gamma^{(2)}/\varepsilon_2}{\frac{\gamma^{(1)}}{\varepsilon_1} + \frac{4\pi \sigma \omega}{c^2} }     & E \mbox{--antisymmetric}.  
\end{array}
\end{eqnarray}  
Complex roots of equations (\ref{modop}) and (\ref{modos})  have been found  by adapting a numerical code based on Newton-Raphson method to complex values. 
\begin{figure}
\centering
\resizebox{0.50\textwidth}{!}
{\includegraphics{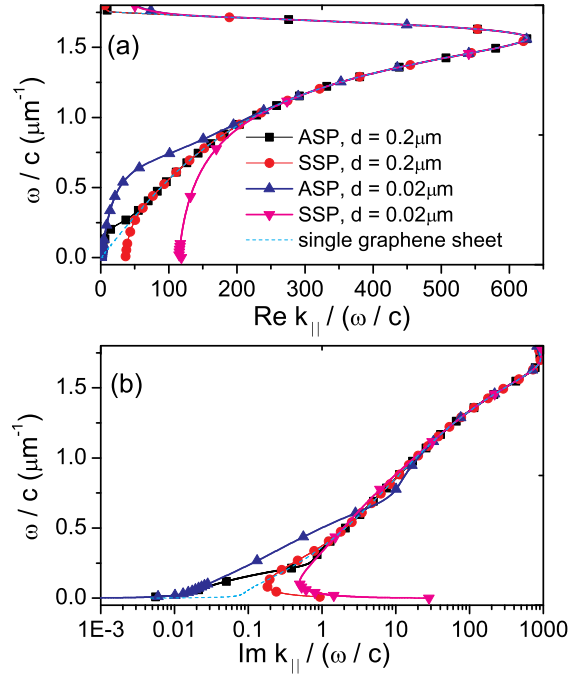}}
\caption{\label{fig:epsart} Dispersion curves for SSP and ASP modes,  calculated for $\mu=0.2$eV, $T = 300$ K, $\gamma_c=0.1$meV, $\varepsilon_1=1$ and $\varepsilon_2=3.9$. (a) $\mbox{Re}\, c k_{||}/\omega$ and (b) $\mbox{Im}\, c k_{||}/\omega$ as a function of $\omega/c$. Plots also show the SP dispersion curve for  a single graphene sheet sandwiched between two dielectric half space with permittivities $\varepsilon_1$ and $\varepsilon_2$.}\label{alphamodop}
\end{figure}
\begin{figure}
\centering
\resizebox{0.50\textwidth}{!}
{\includegraphics{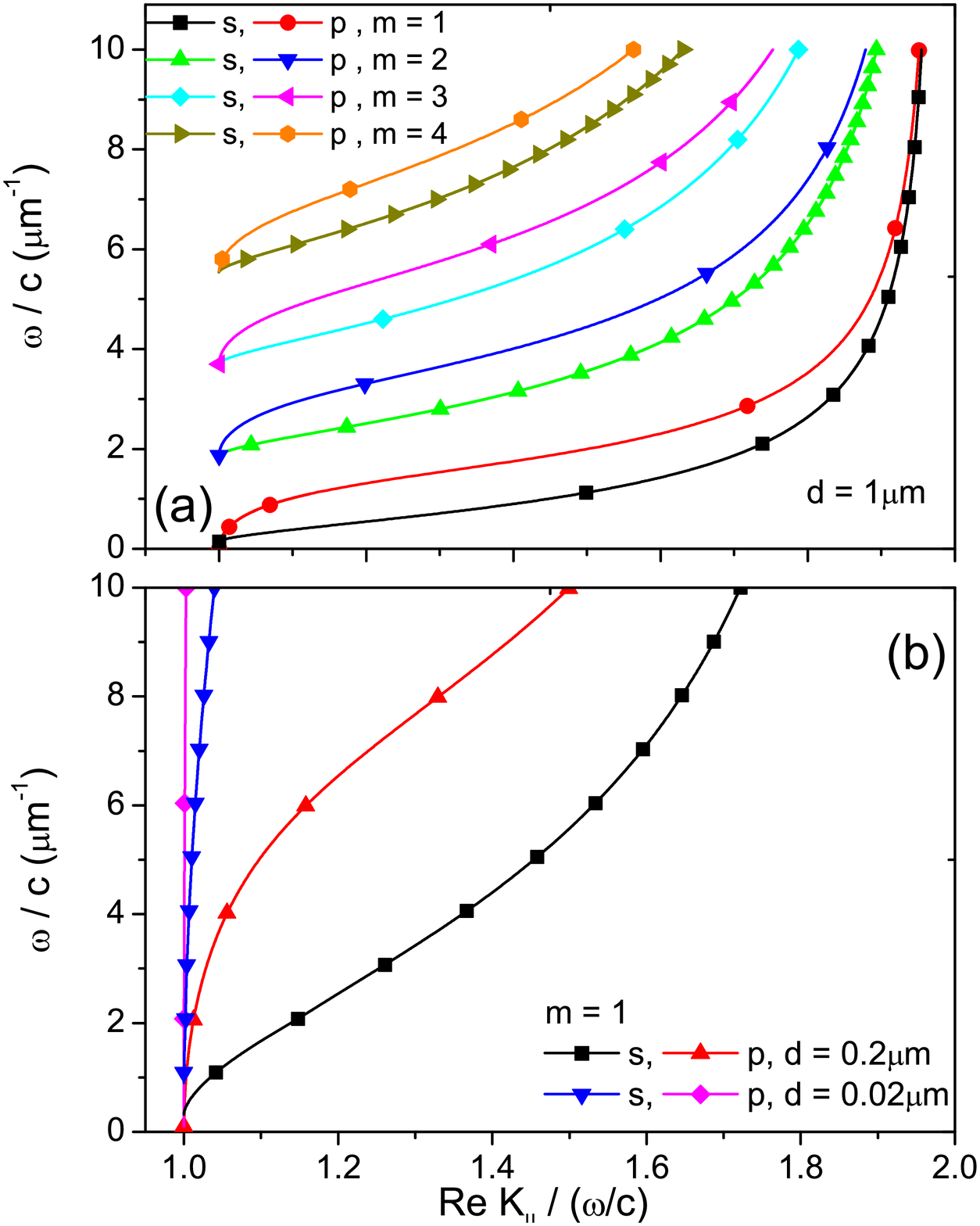}}
\caption{\label{fig:epsart} Real part of the dispersion curves for $s$ and $p$--polarized WG modes  calculated for $\mu=0.2$eV, $T = 300$ K, $\gamma_c=0.1$meV as a function of $\omega/c$. (a) $d=1\,\mu$m and (b) $d=0.2,\,\,0.02\,\mu$m.} \label{alphamodos}
\end{figure}
Figures \ref{alphamodop}a and \ref{alphamodop}b show the real and the imaginary parts of the nondimensional propagation constant $c k_{||}/\omega$  
of SPs as a function of $\omega/c$ obtained by solving Eqs. (\ref{modop})  for $\varepsilon_1=\varepsilon_3 =1$,  $\varepsilon_2 = 3.9$ and for two different waveguide thicknesses, $d=0.2\,\mu$m and $d=0.02\,\mu$m. 
The graphene parameters are $\mu_c=0.2$eV, $\gamma_c=0.1$meV and $T=300$K. 
These figures also show the curves corresponding to the propagation constant of  eigenmodes supported by a single graphene sheet interface (dashed line), \textit{i.e.}, 
a system with the flat graphene sandwiched between two dielectric half space with permittivities $\varepsilon_1$ and $\varepsilon_2$.
Since $\mbox{Im}\,\sigma$ changes sign from positive to negative, due to the presence of the interband term in the conductivity $\sigma$, at $\hbar \omega / \mu_c \approx 1.667$ ($\omega/c \approx 1.667\mu\mbox{m}^{-1}$) Eqs. (\ref{modop}) predicts $p$ polarized surface plasmons restricted to the range below this frequency.   
Moreover,  
the dispersion curves plotted in figure \ref{alphamodop}a exhibit backbending in the vecinity of $\omega/c \approx 1.667\mu\mbox{m}^{-1}$. This behavior has also been observed in the dispersion curves of metallic SPs  when damping 
is taken into account  \cite{greffet1}. At high frequencies, $\omega/c>0.25 \mu\mbox{m}^{-1}$ for $d=0.2\mu$m or $\omega/c>1\mu\mbox{m}^{-1}$ for $d=0.02\mu$m, the  layer between the two graphene sheets is thick relative to the  decay length of SP in medium 2. Therefore,  SPs of the two graphene sheets are essentially uncoupled from each other and their dispersion curves are practically unchanged  from the single interface  case. On the contrary, at lower frequencies, the fields of these modes strongly overlap
through the thin layer (medium 2), leading to  solutions into well separated branches. The upper branch  corresponds to the antisymmetric surface plasmon (ASP) mode and the lower branch corresponds to the symmetric surface plasmon (SSP) mode.

Figure  \ref{alphamodos} shows the real part of the dimensionless propagation constant $c k_{||}/\omega$ of WG  modes as a function of $\omega/c$ obtained by solving the Eqs. (\ref{modop}) and (\ref{modos}) for $p$ and $s$ polarization, respectively,  and for three different waveguide thicknesses, $d=1\,\mu$m (figure  \ref{alphamodos}a) and $d=0.2\,\mbox{and}\,0.02\,\mu$m (figure  \ref{alphamodos}b).  
It has been verified  [not shown in figure \ref{alphamodos}] that the imaginary part of the dimensionless propagation constant of these  modes is less than $10^{-4}$. 
From this figure, it can be seen that the dimensionless propagation constant lies between $n_1=1$ and $n_2=\sqrt{\varepsilon_2} \approx 1.97$ ($n_1<\mbox{Re}\,c k_{||}/\omega<n_2$), thus in the upper (medium 1) and the lower (medium 3)  claddings $\gamma^{(m)}$ ($m=1,\,3$) is almost purely imaginary, $\gamma^{(m)}=i\gamma_{wg}$ where $\gamma_{wg}=\sqrt{k_{||}^2-\left(\frac{\omega}{c}\right)^2}$, and hence,  the field mode  exponentially decays along the $z$ axis. In the core  (medium 2)  $\gamma^{(2)}$ is real and hence the field mode is propagating along the $z$ axis  resulting in a standing wave in this medium.   
This analysis confirms the guided wave nature of these modes, which also exists for a waveguide without graphene,  provided that the core has a higher index of refraction than the cladding,  $\varepsilon_2>\varepsilon_1$.    
Moreover, $\mbox{Re}\,c k_{||}/\omega$  decreases with decreasing values of $d$, thus the WG mode is less tightly bound as the spacing between the two graphene sheets decreases.   
This fact highlights  the small degree of localization of WG  modes for very small thicknesses. 
 
Once the zeros of Eqs. (\ref{modop}) and (\ref{modos})  are determined, the contribution of each pole to the total decay rate has been calculated by the residues method \cite{fordyweber}. These contributions dominate the behavior of the spontaneous emission on frequency regions where the  eigenmodes are well defined. 

In order to obtain an approximated analytic expression for the normal decay rate into each of the eigenmodes, each pole  contribution is extracted of Eq. (\ref{PP}) in the small losses limit for which the imaginary part of the eigenmode  propagation constant can be neglected,     
\begin{eqnarray}\label{decaimiento_normal}
\left[\frac{P}{P_0}\right]_{\bot,mod}= 
\frac{3}{2}\,  \pi \left(\frac{k_{mod}}{k_1}\right)^3 \frac{e^{-\gamma_{mod}2(z^{'}-d)}}{\gamma_{mod}}\,\mbox{Res}\,r_p^{(1,3)},
\end{eqnarray}
where $k_{mod}$ is the real part of the propagation constant of a particular eigenmode, $k_{spp}$ for SP  or $k_{wg}$ for WG modes (both quantities higher than the modulus of the photon wave vector in vacuum $k_0=\omega/c$), $\gamma_{mod}=\frac{\gamma^{(1)}}{i}=\sqrt{k_{mod}^2-(\frac{\omega}{c})^2}$ is the $z$ component of the wave vector in medium 1, and Res is the residue of the integrand in (\ref{PP}) at the pole $k_{||}=k_{mod}$,
\begin{eqnarray}
\mbox{Res}\,r_p^{(1,3)}=\lim_{k_{||} \to k_{mod}} (k_{||} - k_{mod}) \, r_p^{(1,3)}.
\end{eqnarray}
Similarly, the contribution of each eigenmodes to the parallel  decay rate  can be approximated by evaluating each pole contribution, in the small losses limit, in Eq. (\ref{PP_paralelo}),
%
\begin{eqnarray}\label{decaimiento_paralelo}
\left[\frac{P}{P_0}\right]_{||,mod}= \frac{3}{4}\,  \pi 
\left(\frac{k_{mod}}{k_1}\right) \frac{e^{-\gamma_{mod}2(z^{'}-d)}}{\gamma_{mod}}\,\mbox{Res}\,r_s^{(1,3)}\nonumber\\
 + \frac{3}{4}\,  \pi  \left(\frac{k_{mod}}{k_1}\right) \frac{\gamma_{mod}\,e^{-\gamma_{mod}2(z^{'}-d)}}{k_1^2}\,\mbox{Res}\,r_p^{(1,3)},
\end{eqnarray}
%
where 
\begin{eqnarray}
\mbox{Res}\,r_s^{(1,3)}=\lim_{k_{||} \to k_{mod}} (k_{||} - k_{mod}) \, r_s^{(1,3)}.
\end{eqnarray}
Note that, for a horizontal dipole, there are  $s$ and $p$ polarized decay channels involved  in the first and in the second term in Eq. (\ref{decaimiento_paralelo}), respectively. Since only $p$ polarized SPs exist, only the second term  corresponds to the decay rates into SPs. 

From figure \ref{alphamodop}a, it is clear that the modulus of the photon wave vector is negligible compared with the propagation constant of SPs ($n_m\,\omega/c <<k_{spp}$, where $m=1,\,2,\,3$ and $k_{spp}$ denotes either the symmetric or the antisymmetric SP propagation constant). 
As a consequence, $\gamma_{spp} \approx k_{spp}$ and hence, the pole contribution in Eq. (\ref{decaimiento_paralelo}) corresponding to the plasmonic contribution (SSP or ASP) can be approximated as follows 
\begin{eqnarray}\label{paralelo_perpendicular}
\left[\frac{P}{P_0}\right]_{||,spp} &  = & \frac{3}{4}\,  \pi  \left(\frac{k_{spp}}{k_1}\right) \frac{\gamma_{spp}\,e^{-\gamma_{spp}2(z^{'}-d)}}{k_1^2}\times \nonumber\\
\,\mbox{Res}\,r_p^{(1,3)} & \approx &	 \frac{3}{4}\,  \pi  \left(\frac{k_{spp}}{k_1}\right)^3 \,\frac{e^{-\gamma_{spp}2(z^{'}-d)}}{\gamma_{spp}}\times \nonumber\\
\,\mbox{Res}\,r_p^{(1,3)} & = & \frac{1}{2}\left[\frac{P}{P_0}\right]_{\bot,spp}, 
\end{eqnarray}
where in the last equality  Eq. (\ref{decaimiento_normal}) has been used. From  Eq. (\ref{paralelo_perpendicular}) we see that the spontaneous decay rate into SPs of a single emitter whose dipole moment is perpendicular to the graphene monolayers  is  twice the corresponding value to  the same emitter but with the dipole moment in the  parallel direction to the graphene monolayers.

\section{Results and discussion} \label{resultados}

Initially, we  analyze the contribution of eigenmodes to the total decay rate of a  dipole located at distance $l=z^{'}-d=0.01\mu$m from the surface of the graphene waveguide. Thickness $d=1\mu$m,  constitutive parameters of the dielectric slab and graphene sheets are $\varepsilon_1=1$, $\varepsilon_2=3.9$, and  $\mu_c=0.2$eV, $\gamma_c=0.1$meV, $T=300$K, respectively.   

\begin{figure}
\centering
\resizebox{0.60\textwidth}{!}
{\includegraphics{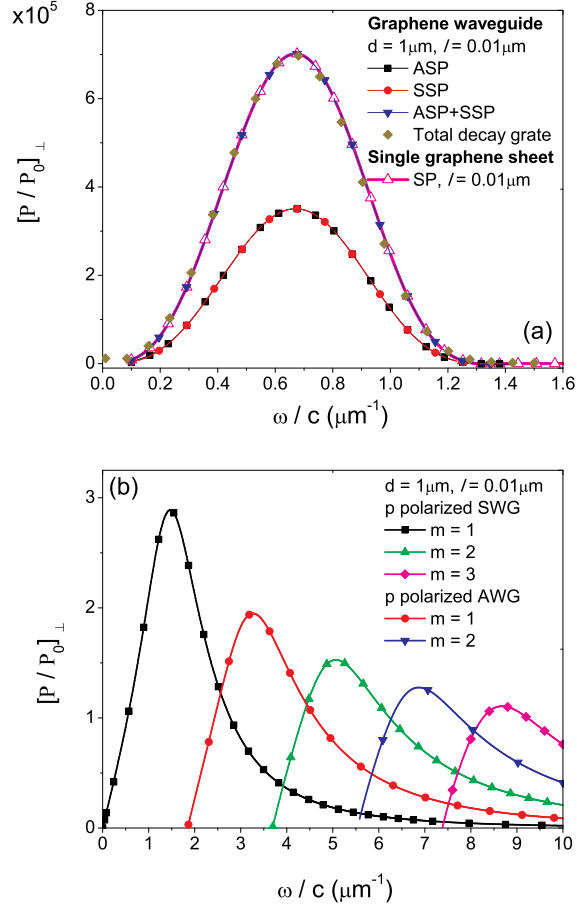}}
\caption{\label{fig:epsart} 
(a) Total decay rate and SP contributions to the decay rate of a vertical dipole placed at a distance $l = 0.01\mu$m above a graphene waveguide as a function of $\omega/c$. (b) WG mode contributions to the decay rate as a function of $\omega /c$.  The  waveguide parameters are $\mu_c=0.2$eV, $\gamma_c=0.1$meV, $\varepsilon_1=\varepsilon_3=1$, $\varepsilon_2=3.9$, $d=1 \mu$m and $T=300$K. Plot (a) also shows a curve corresponding to the SP contribution of a vertical dipole placed at a distance $l = 0.01\mu$m above a single graphene sheet.
%
}\label{superficieplana}
\end{figure}

\begin{figure}
\centering
\resizebox{0.60\textwidth}{!}
{\includegraphics{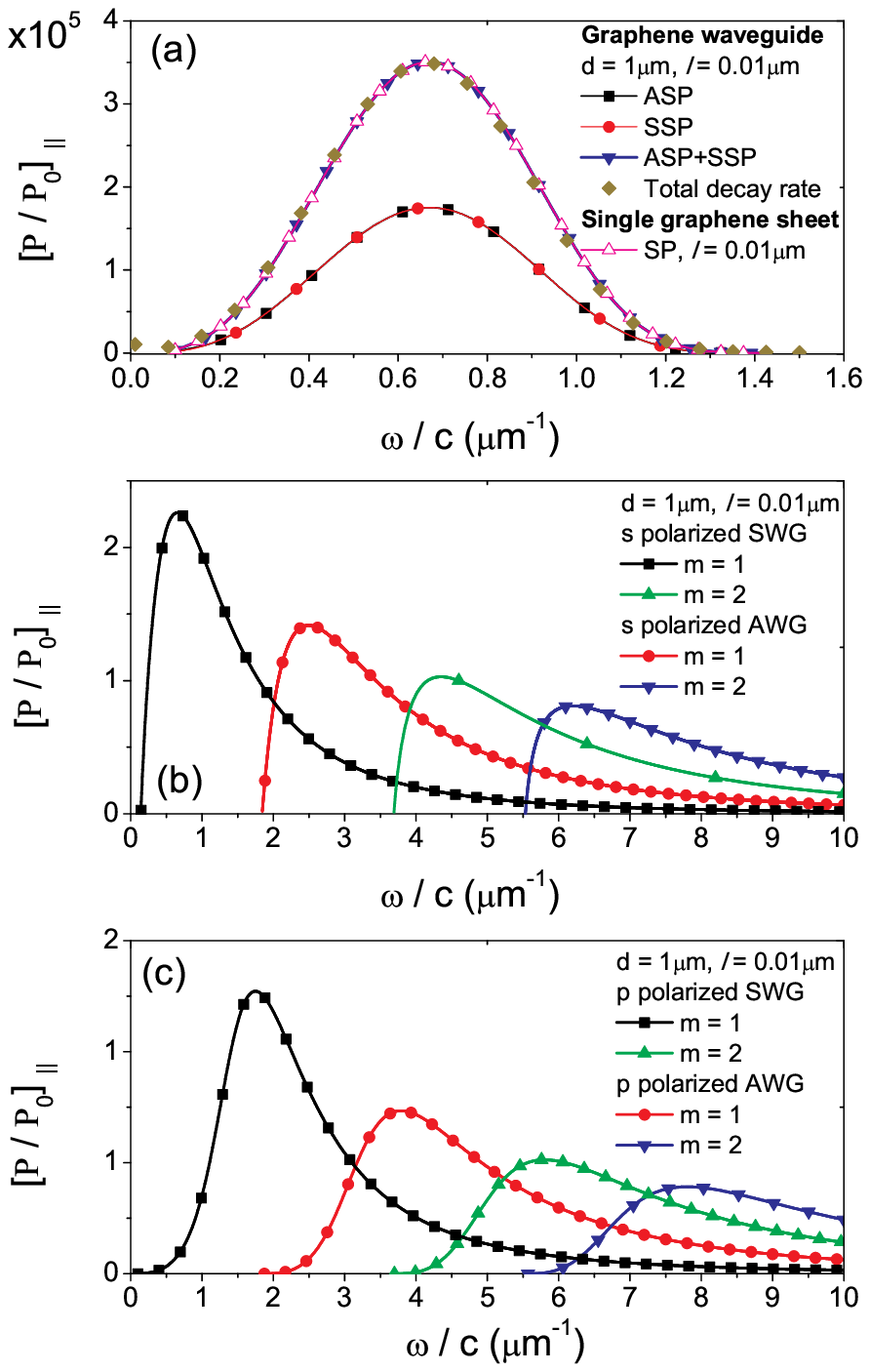}}
\caption{\label{fig:epsart} 
(a) Total decay rate and SP contributions to the decay rate of a parallel dipole placed at a distance $l = 0.01\mu$m above a graphene waveguide as a function of $\omega/c$. $s$ polarized  (b) and $p$ polarized (c) WG mode contributions to the decay rate as a function of $\omega /c$. Plot (a) also shows a curve corresponding to the SP contribution of a parallel dipole placed at a distance $l = 0.01\mu$m above a single graphene sheet.
%
The  waveguide parameters are 
the same  as in figure \ref{superficieplana}.}\label{panel3}
\end{figure}

\begin{figure}
\centering
\resizebox{0.50\textwidth}{!}
{\includegraphics{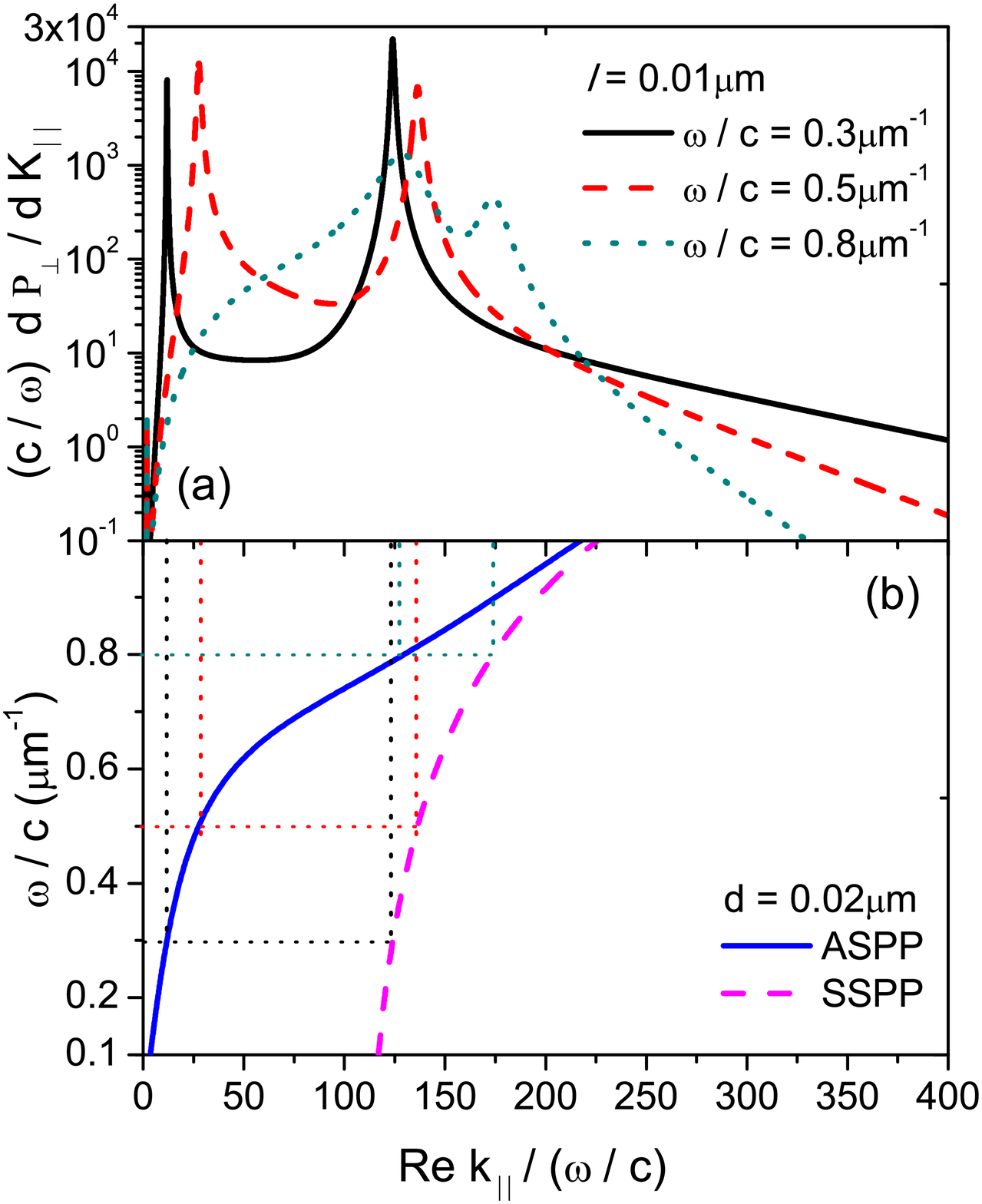}}
\caption{\label{fig:epsart} (a) $k_{||}$ space power spectrum for a vertical dipole placed a distance $l=0.01\mu$m  above a graphene  waveguide and for frequencies $\omega/c=0.3,\,0.5$ and $0.8\mu\mbox{m}^{-1}$. 
 (b) Dispersion curves for $p$ polarized SPs. The  waveguide parameters are $\mu_c=0.2$eV, $\gamma_c=0.1$meV, $\varepsilon_1=\varepsilon_3=1$, $\varepsilon_2=3.9$, $d=0.02 \mu$m and $T=300$K.}\label{dP_dK}
\end{figure}

Figure \ref {superficieplana} and figure \ref {panel3} show the normal  and the parallel decay rates as a function of $\omega/c$ frequency, respectively. 
The total decay rate is the numerical result of Eqs. (\ref{PP})  and (\ref{PP_paralelo}) for perpendicular and parallel dipole orientation, respectively. 
In the former case only $p$ polarized eigenmodes can be excited, while in the second case $p$ and $s$ polarized eigenmodes can be excited. 
These figures also show the different contributions to the spontaneous emission  rate obtained by using Eqs. (\ref{decaimiento_normal}) and (\ref{decaimiento_paralelo}).  
The acronym SWG  refers to symmetric WG modes, \textit{i.e}, to solutions of the first Eq. (\ref{modop}) or to solutions of the first Eq. (\ref{modos}). Similarly, the acronym AWG refers to antisymmetric WG modes, \textit{i.e}, to solutions of the second Eq. (\ref{modop})  or to solutions of the second Eq. (\ref{modos}). 
From these figures, it is clear that the decay rates through the WG modes  is much lower (by  a factor $10^5$) than the corresponding decays through the excitation of SSP and ASP modes. This is true because  the field of SPs  concentrates near the surface much more strongly than  the field of the WG modes. 
As Figures \ref {superficieplana}a and \ref {panel3}a show,  the total decay rate and the sum of the contributions from SP modes  overlap. 
%
These figures   
also shows the curve corresponding to the same emitter placed at a distance $l=0.01\mu$m above a single graphene  sheet separating medium 1 (where the dipole is placed) from medium 2. We see that the contribution of both SSP  and ASP modes  coincide in the whole frequency range, in agreement with the fact that,     
for sufficiently large $d$, both the symmetric and the antisymmetric branches merge into the dispersion curve of the single SP mode supported by a graphene monolayer (dashed line in fig. \ref{alphamodop}). 
Moreover, the total contribution to the decay rate of SPs (ASP $+$ SSP) agree with the corresponding contribution to the decay rate of SPs on a single graphene monolayer, according to the fact that for large enough $d$ values and for low values of    
$l$ ($l<<d$) the system formed by the source and the graphene waveguide  resembles a system formed by the source and a single graphene sheet. 
As the thickness $d$ is decreased, a splitting of the dispersion curves into the symmetric and the antisymmetric mode curves occurs (figure \ref{alphamodop}a) and, as a consequence  different contributions to the decay rate by the symmetric and the antisymmetric SPs are expected.
This fact can be seen in 
figure \ref{dP_dK}a, where we have plotted  the integrand of Eq. (\ref{PP}), $(c/\omega)\, d P_{\bot}/ d k_{||}$ ($k_{||}$--\textit{space power spectrum}), as a function of the dimensionless parallel  wave vector $c k_{||}/\omega$ for a vertical dipole placed at distance $l=0.01\mu$m from the waveguide and   for several  $\omega/c=0.3,\,0.5,\,\mbox{and }0.8\mu$m$^{-1}$ frequency values. 
All curves show two prominent peaks due to excitation of antisymmetric (low wavenumber) and symmetric (high wavenumber) SPs.  The correspondence between these peaks and the SP resonances of the graphene waveguide is evidenced in figure \ref{dP_dK}b, where we have plotted the real part of the dispersion curves for symmetric and antisymmetric SP modes. Moreover, as the frequency increases both peaks are widened in accordance with the fact that, in this frequency range,  the imaginary part of the dimensionless propagation constant of both symmetric and antisymmetric SPs increases as the frequency increases (figure \ref{alphamodop}b). 
Similar behavior has been observed [not shown in fig. \ref{dP_dK}] in the case for which the dipole is placed  parallel to the graphene waveguide.    
%
\begin{figure}
\centering
\resizebox{1.0\textwidth}{!}
{\includegraphics{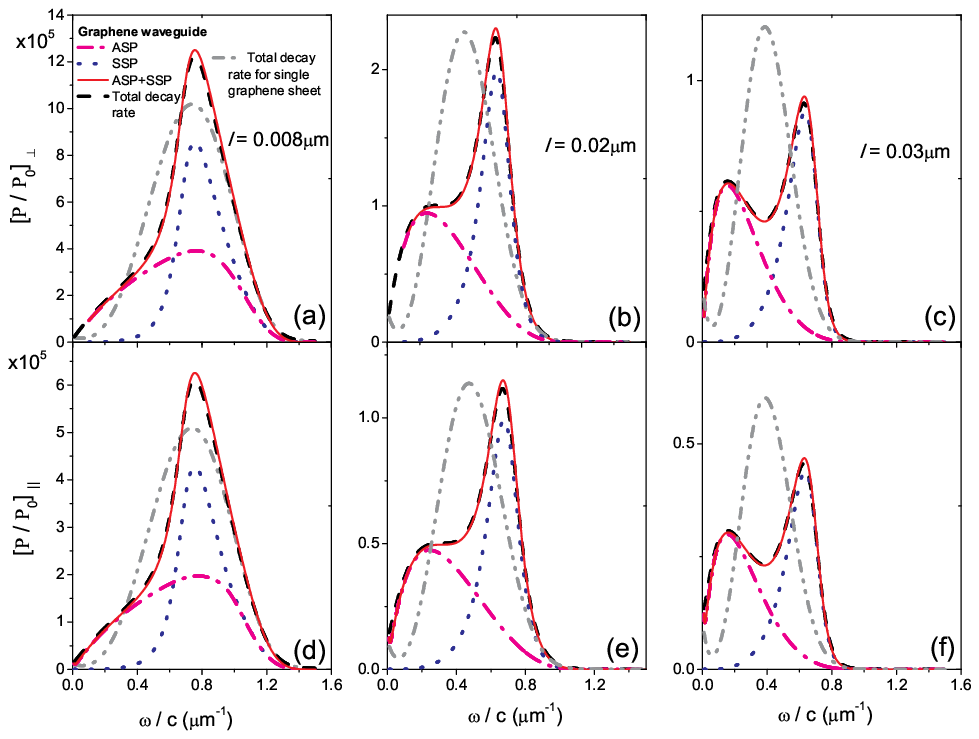}}
\caption{\label{fig:epsart} 
Total decay rate and different contributions to the decay rate as a function of $\omega/ c$ of a dipole placed  above a graphene waveguide for the same parameters as in figure \ref{dP_dK}. Plots also show a curve corresponding to total decay rate of a same dipole placed above a single graphene sheet. The orientation of the dipole is vertical for the three panels on the top row and it is horizontal for the three panels on the bottom row. 
The distance $l=0.008\mu$m (a) and (d), $l=\,0.02\mu$m (b) and (e), and $l=\,0.03\mu$m (c) and (f).}
\label{SE1}
\end{figure}
%
From figure \ref{dP_dK}a, we observe that most of the contribution to the power spectrum is given by SP peaks, and observing this figure allows us  to assert that each of these peaks contributes with a different value to the spontaneous decay rate. This fact can be viewed in figure \ref{SE1} in which the frequency dependence of 
the total decay rate and the SP contributions 
are plotted   
for the same parameters as in figure \ref{dP_dK}, except now for distances $l = 0.008,\,0.02,$ and $\,0.03\mu$m. 
From this figure, it can be seen that the differences between the total decay rate  and the sum of the contributions from SP modes are small.    
On the other hand, for small values of $l$,  the level of the total decay rate for graphene waveguide is larger than the corresponding level reached for a single graphene sheet (figs. \ref{SE1}a and \ref{SE1}d). As the distance $l$ is increased, the level of the total decay rate for graphene waveguide is equal or even lower than for a single graphene sheet (figs. \ref{SE1}b, \ref{SE1}c and \ref{SE1}e, \ref{SE1}f). 

It is worth noting that the shape of the spontaneous emission spectrum is strongly influenced by the SP branches. For instance, when the distance  $l$ is small, 
the coupling strength between the  near field emitted by the source  and SPs with large wave vectors is high, leading to  an increase of the decay rate in the corresponding high frequency range. 
In this range, the dispersion of the ASP and SSP modes are similar to that of the SP mode on a single graphene interface (figure \ref{alphamodop}a) and, thus, the frequency of the peak in the total decay rate curve for graphene waveguide coincides with the frequency of the peak corresponding to a single graphene sheet (figures \ref{SE1}a and \ref{SE1}d).
%
 As  $l$  distance is increased, the strength in this range of frequency falls  because the  near field  can only excite SPs with  increasingly smaller wave vectors.
In this range, the fields of the ASP and SSP modes strongly overlap through the thin layer, leading to well separated branches (figure \ref{alphamodop}a), where the upper branch corresponds to the ASP mode and the lower branch corresponds to the SSP mode. As a consequence, the frequency of the peak in the ASP decay rate curve is larger than that  corresponding to decay rate through the excitation of SSP modes, 
explaining the separation between  the peaks in the decay rate curves for graphene waveguide and for a single  graphene sheet observed in figures \ref{SE1}b, \ref{SE1}c, \ref{SE1}e, and \ref{SE1}f.
Moreover,  as $l$ is increased, the peak of the curves  shifts to lower frequencies and, for  large enough $l$ values the total decay rate for graphene waveguide exhibits a double peak structure, as can be seen in figures \ref{SE1}c and \ref{SE1}f.
%
Note that, according to Eq. (\ref{paralelo_perpendicular}), the values of the decay rates plotted in figures  \ref{SE1}a, \ref{SE1}b and \ref{SE1}c are approximately twice the values of the decay rates plotted in figures \ref{SE1}d, \ref{SE1}e and \ref{SE1}f, respectively.

It is well established nowadays that the phenomenon of the  spontaneous emission can be understood in the framework of quantum electrodynamics. In the weak coupling regime, within the   
dipole approximation, the decay constant for a radiating dipole located at $\vec{x^{'}}=z^{'}\hat{z}$ is given by Fermi's golden rule  (see Refs \cite{yablonovich,brongersma}), 
\begin{eqnarray}\label{fermi}
\frac{1}{\tau}=\frac{2\pi}{\hbar^2}\,|\vec{p}\cdot \alpha \vec{E}(z^{'},\omega)|^2 D_{\mbox{\tiny{2D}}}(\omega), 
\end{eqnarray}
where $\vec{p}$ is the dipole moment matrix element, $D_{\mbox{\tiny{2D}}}$ is  the surface plasmon density of states and $\alpha$  is a normalization factor related to the  vacuum fluctuation energy,
\begin{eqnarray}\label{alpha}
|\alpha|^2=
\frac{ \hbar \omega / 2}{\frac{S}{8\pi} \int_{-\infty}^{\infty} \left\{\varepsilon(z) |\vec{E(z,\omega)}|^2+|\vec{H(z,\omega)}|^2\right\} dz}  ,
\end{eqnarray}
If  the dipole lies in the $x-y$ plane ($\vec{p}=p_x\hat{x}+p_y\hat{y}$), so that $p_x^2=p_y^2=p^2/2$, then the  decay rate (\ref{fermi}) can be written as 
\begin{eqnarray}\label{P_paralelo}
\frac{1}{\tau_{||}}=\frac{ \pi^2 \omega |\vec{p}|^2}{\hbar V_{\mbox{\tiny{eff}}}(z^{'},\omega)} 
\end{eqnarray}
where $S$ is the in--plane quantization area, $V_{\mbox{\tiny{eff}}}=S L_{\mbox{\tiny{eff}},||}$ is the effective mode volume, and $L_{\mbox{\tiny{eff}},||}$ is the effective mode length in $z$--axis direction for a dipole oriented parallel to the $x-y$ plane
\begin{eqnarray}\label{L_paralelo}
L_{\mbox{\tiny{eff}},||}(z^{'},\omega)=\frac{\frac{1}{2}\int_{-\infty}^{\infty} \left\{\varepsilon(z) |\vec{E(z,\omega)}|^2+|\vec{H(z,\omega)}|^2\right\}
dz }{|\vec{E}_{||}(z^{'},\omega)|^2} 
\end{eqnarray}
with $\vec{E}_{||}$ being the parallel  component of the plasmon field. 
The surface plasmon density of states is obtained calculating  the number of corresponding modes in the two--dimensional $k$ space,
\begin{eqnarray}\label{LDOS}
D_{\mbox{\tiny{2D}}}=\frac{S}{2\pi} k \frac{dk}{d\omega}= 
\frac{S\,\omega}{2\pi v_p(\omega) v_g(\omega)},
\end{eqnarray}
with $v_p$ and $v_g$ representing the phase and group velocities of the SP mode calculated at the dipole emission frequency, respectively. 
Inserting Eq. (\ref{LDOS}) into Eq. (\ref{P_paralelo}), using Eq. (\ref{L_paralelo}), we find the normalized 
emission lifetime to be
\begin{eqnarray}\label{fermi2_paralelo}
\frac{\tau_0}{\tau_{||}}=\frac{3}{8}\frac{c^2\,\lambda}{v_p v_g L_{\mbox{\tiny{eff}},||}(z^{'},\omega)} 
\end{eqnarray}
where $\lambda=2\pi c/\omega$ is the wavelength of the source, $\tau_0^{-1}=4 |\vec{p}|^2 \omega^3/(3 \hbar c^3)$  is the spontaneous emission decay constant for vacuum derived from cavity quantum electrodynamics considerations \cite{glauber}. 

On the other hand, if  the dipole is oriented in the $z$ axis, $\vec{p}=p\hat{z}$,  following the same procedure as used to deduce Eq. (\ref{fermi2_paralelo}), the  decay rate (\ref{fermi}) can be written as 
\begin{eqnarray}\label{fermi2_perpendicular}
\frac{\tau_0}{\tau_{\bot}}=\frac{3}{4}\frac{c^2\,\lambda}{v_p v_g L_{\mbox{\tiny{eff}}}(z^{'},\omega)} 
\end{eqnarray}
where
\begin{eqnarray}\label{L_perpendicular}
L_{\mbox{\tiny{eff}},\bot}(z^{'},\omega)=\frac{\frac{1}{2}\int_{-\infty}^{\infty} \left\{\varepsilon(z) |\vec{E(z,\omega)}|^2+|\vec{H(z,\omega)}|^2\right\}
dz }{|\vec{E}_{\bot}(z^{'},\omega)|^2} 
\end{eqnarray}
is the effective mode length  when the dipole is oriented in $z$--axis direction and $\vec{E}_{\bot}$ is the perpendicular  component of the plasmon field.

It is known that \cite{novotny} if we identify the dipole matrix element $\vec{p}$ in Eqs. (\ref{fermi2_paralelo}) and (\ref{fermi2_perpendicular}) with the classical dipole in Eqs. (\ref{PP})  and (\ref{PP_paralelo}), then the normalized decay constant is equal to the normalized classical radiation power, \textit{i. e.}, $\tau_0/\tau=P/P_0$ for both dipole  orientations, parallel and perpendicular to the $x-y$ plane.  
In this framework, the enhancement of the spontaneous emission rate can be quantified by the reduction of effective mode length and the group velocity of surface plasmons, \textit{i.e.},  $L_{\mbox{\tiny{eff}}}\,v_g$.   
	
To illustrate, we explore the tunability  of the spontaneous emission  by varying the chemical potential $\mu_c$ on graphene sheets.  
As in the classical treatment, from the quantum point of view the decay rate of a dipole with  parallel orientation is close to half of the corresponding decay of the same dipole with  perpendicular orientation in the whole frequency range where SP  modes exist.
This fact can be viewed as follows.  Due to the fact that the modulus of the photon wave vector is negligible compared with the propagation constant of SPs, $\omega/c<<k_{spp}$, the modulus of both components of SP electric fields, $|\vec{E}_{||}|$ and $|\vec{E}_{\bot}|$,  are approximately equal. Therefore, from Eqs. (\ref{fermi2_paralelo}) and (\ref{fermi2_perpendicular}) it follows that $[P/P_0]_{||}\approx  [P/P_0]_{\bot} / 2$. We have numerically verified this assertion. Thus, we only show examples corresponding to an emitter whose dipole moment is perpendicular to graphene  monolayers.   The  waveguide parameters chosen are the same  as in figure \ref{SE1}.   
\begin{figure}[htbp!]
\centering
\resizebox{0.70\textwidth}{!} 
{\includegraphics{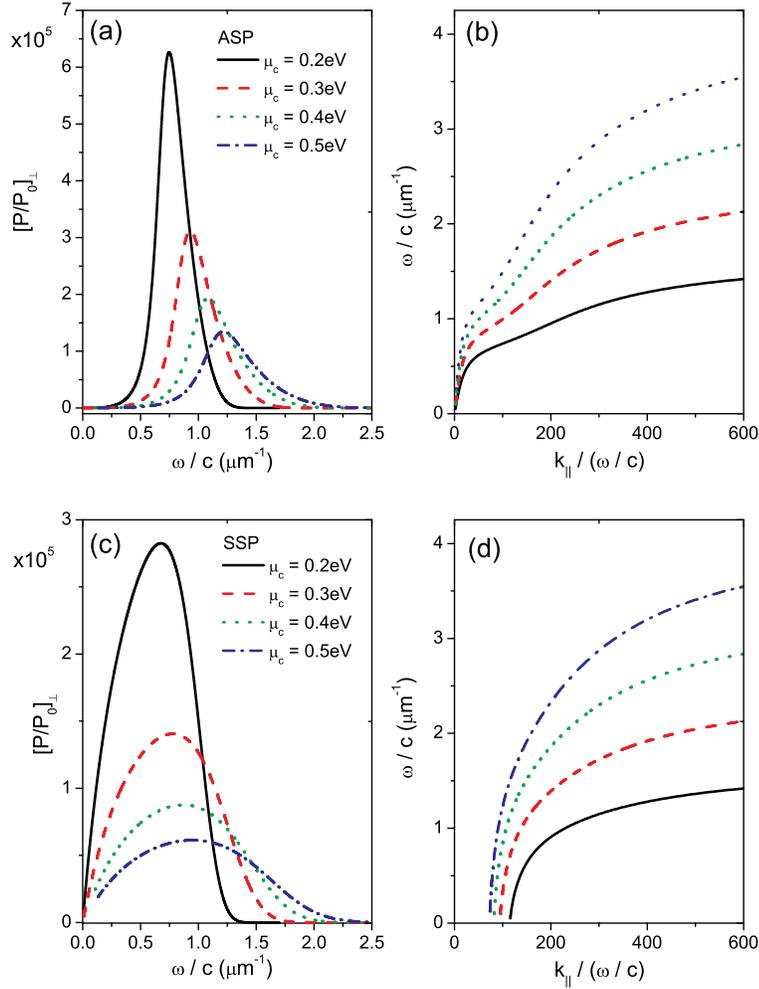}} 
\caption{\label{fig:epsart} Decay rate of a vertical dipole located at $l=0.01\mu$m from the graphene waveguide into (a) ASP and (c) SSP  for different values of the chemical potential $\mu_c$ (0.2, 0.3, 0.4 and 0.5eV). Dispersion curves for (b) ASPs  and (d) SSPs calculated for the same chemical potential values of (a) and (c). The  waveguide parameters are $\gamma_c=0.1$meV, $\varepsilon_1=\varepsilon_3=1$, $\varepsilon_2=3.9$, $d=0.02 \mu$m and $T=300$K.}\label{panel_fermi_1}
\end{figure}

Figure \ref{panel_fermi_1}a and \ref{panel_fermi_1}c  show the normalized decay rate of a  vertical dipole placed at a distance $l=0.01\mu$m above the graphene waveguide into the symmetric and the antisymmetric SPs, respectively, calculated by using Eq. (\ref{fermi2_perpendicular}). The drop in these curves occur at the frequency where $\sigma$ changes sign from positive to negative, $\hbar\omega\approx1.667 \mu_c$. At this frequency the slope of the dispersion curves  tends to zero, as can be seen in Figures  \ref{panel_fermi_1}b and \ref{panel_fermi_1}d where these curves have been plotted  for symmetric and antisymmetric SPs, respectively. Unlike section \ref{plasmons}, the dispersion curves shown in figures \ref{panel_fermi_1}b and \ref{panel_fermi_1}d  have been calculated neglecting losses in graphene sheets, a basic approach to derive Eqs. (\ref{fermi2_paralelo}) and  (\ref{fermi2_perpendicular}) from   the  quantization scheme 
of SP field  \cite{greffet2}. 
We have verified that results obtained from Eq. (\ref{fermi2_perpendicular}) are in agreement with those obtained from  classical formalism by using Eq. (\ref{decaimiento_normal}). 
%
%
\begin{figure}[htbp!]
\centering
\resizebox{0.70\textwidth}{!}
{\includegraphics{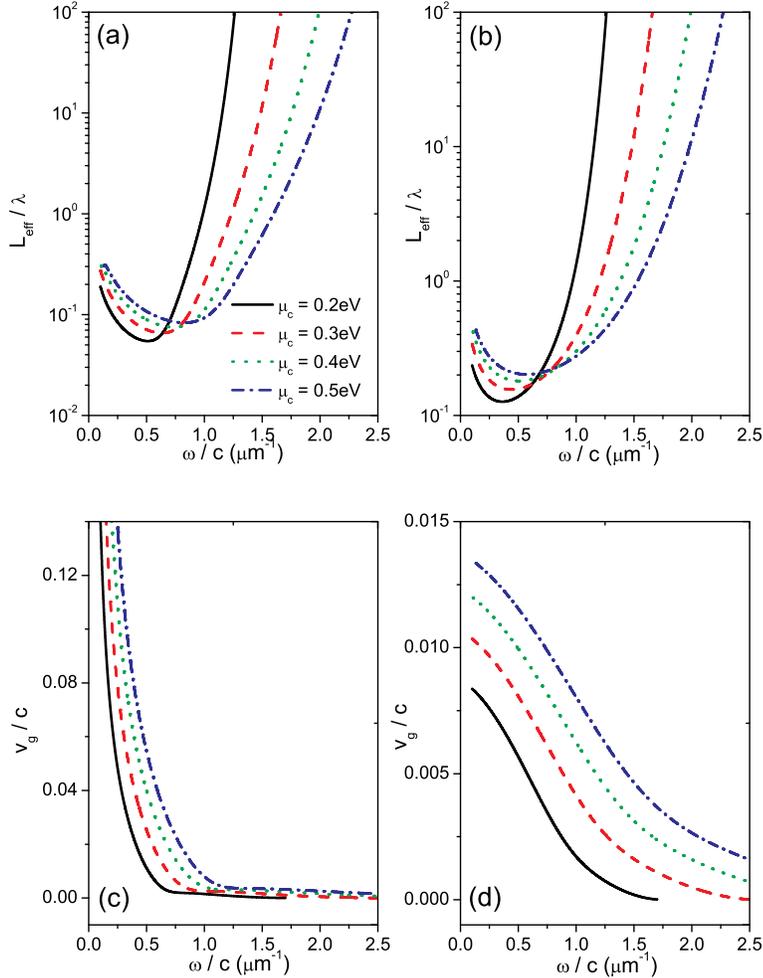}} 
\caption{\label{fig:epsart} Normalized mode length for (a) ASP and (b) SSP. Normalized group velocity for (c) ASPs and (d)  SSPs. The curves have been calculated for different values of the chemical potential $\mu_c$ (0.2, 0.3, 0.4 and 0.5eV). The vertical dipole is located at $l=0.01\mu$m from the graphene waveguide. The  waveguide parameters are $\gamma_c=0.1$meV, $\varepsilon_1=\varepsilon_3=1$, $\varepsilon_2=3.9$, $d=0.02 \mu$m and $T=300$K.}\label{panel_fermi_2}
\end{figure}

Figure \ref{panel_fermi_1}a and figure \ref{panel_fermi_1}c show that the decay rate peak  shifts to blue, as the value of $\mu_c$ increases. This behavior can be understood with the help of  figure \ref{panel_fermi_2}, where  both the effective mode length $L_{\mbox{\tiny{eff}}}$ and the group velocity of surface plasmons as a function of $\omega/c$ have been plotted. From figures \ref{panel_fermi_2}a and \ref{panel_fermi_2}b, it can be seen  that the  curves of  $L_{\mbox{\tiny{eff}}}$  exhibit a minimum at a frequency value slightly lower than the frequency where the curves corresponding to spontaneous decay rates shown in figures \ref{panel_fermi_1}a and \ref{panel_fermi_1}b exhibit a maximum. The reduction of the group velocity with the frequency, shown in figures \ref{panel_fermi_2}c and \ref{panel_fermi_2}d, moves the minimum of the denominator in Eq. (\ref{fermi2_perpendicular}) ($L_{\mbox{\tiny{eff}}}\,v_g$) toward the position of the spontaneous decay rate peak.

\section{Conclusions} \label{conclusiones}

We have presented an exhaustive study of the spontaneous emission rate of a single emitter (atom or molecule) in a planar graphene waveguide formed by two parallel graphene monolayers with an insulator spacer layer. We developed an analytical classical method and obtained a rigorous solution in a closed integral form. 
This solution has the functional form  corresponding to a dielectric or a metallic slab,  although the current density induced in the graphene sheets leads to a marked difference between the reflection coefficient corresponding to a graphene waveguide  and a  waveguide without graphene monolayers (bare waveguide).

We  separately calculated the contribution of symmetric and antisymmetric eigenmodes -- SPs and  WG modes-- to the total decay rate. In the presented examples, we have varied the location of the emitter for both dipole moment orientations, parallel and perpendicular to the graphene monolayers. 
The dipole moment perpendicular to the  graphene monolayers cannot couple to $s$--polarized eigenmodes,  whereas the dipole moment parallel to the  graphene monolayers couples to both $s$ and $p$ eigenmodes. 
The emphasis has been centered around the plasmonic channels, since their contributions  play a prominent role  in the  spontaneous emission rate of single emitters placed  near the graphene waveguide.  
An interesting result revealed in this study is related with a redistribution of the emitted power by a dipole located near a waveguide structure, \textit{i.e.}, the influence of the SP branches on the shape of the emission spectrum. We have shown that by increasing the distance between the emitter and the graphene waveguide, one can obtain spectral behaviors ranging from a single peak curve similar to that of a single graphene sheet  to a double peak curve.    

The coupling efficiency between the emitter and SP modes was also  studied from an equation based on Fermi golden rule. Our examples show that the reduction of both the effective mode length and the group velocity of surface plasmons lead to an enhancement of the spontaneous emission rate. We have shown that, by tuning of the chemical potential of graphene, it is possible to modify the density of states as well as the effective mode volume of SPs leading to unprecedented control over the location and magnitude of the spontaneous emission rate.

The possibility to vary the chemical potential of one of the graphene sheets with respect to  the other one fixed,  allows another degree of freedom to modify  SP branches, with fields no longer symmetric or antisymmetric across the gap dielectric layer, and their influence on the spontaneous emission rate. Although we are planning to report the results of such study in a future paper, as a first step, here we have restricted ourselves to performing an analysis of the symmetric waveguide in which the  two conductivities of the graphene sheets are equal. 

\section*{Acknowledgment}
The author acknowledge the financial support of Consejo Nacional de Investigaciones Cient\'{\i}ficas y T\'ecnicas, (CONICET, PIP 451).

\end{document}